\documentclass[12pt, prd, showpacs]{revtex4}
\input{tcilatex}

\begin{document}

\title{Ettore Majorana: quantum mechanics of destiny}
\author{O. B. Zaslavskii}
\affiliation{Department of Mechanics and Mathematics, Kharkov V.N.Karazin National
University, \\
Svoboda Square 4, Kharkov 61077, Ukraine}
\email{ozaslav@kharkov.ua}

\begin{abstract}
We show that a number of apparently separate extravagant Majorana's actions,
including those connected with his disappearance are united by the invariant
of behavior. It is based on "carrying into life" the principles of quantum
theory. We argue that the underlying motive force in the story of his
disappearance consisted in existential intention to overcome the fixed frame
of personal identity and dichotomy "life -- death" and mimic breakthrough to
plurality of worlds.
\end{abstract}

\maketitle

% It is always \today, today, but any date may be explicitly specified


\section{Introduction}

In history of science Ettore Majorana (1906 - ?) is singled out not only by
his outstanding works on quantum theory (see, for example, a survey on
scientific heritage of Majorana in \cite{rec}) but also by his unusually
striking personality. Unfortunately, there is only a few works devoted
specifically to Majorana. Mainly, they are published in Italian -- see, for
example, brief bibliography in a recent paper \cite{step}. It comes without
surprise that it attracts attention of not only historians of science but
also of peoples of art. In relatively recent years, it stimulated a series
of artistic (in particular, theatre) works \cite{art}. Probably, it is
partially connected with the fact that in situations where a historian can
say nothing, he hands on the baton to an artist, analysis being replaced by
fantasy. Meanwhile, in spite of all uncommonness (sometimes even
extravagancy) of what is known about behavior and actions of Majorana, in
our view we can advance (at least partially) in understanding the logic that
drove apparently extravagant actions of this exceptional man. We will see
that in some these actions there are common features and this reveals very
unusual character of what is traditionally designated as unity between life
and work of a scientist. Their analysis leads to a rather unexpected
conclusion and gives reason to speak about (using terminology of modern
semiotics) "text of behavior". The aim of the present work consists just in
uncovering such a text and its analysis. (As far as Majorana's scientific
activity is concerned, we touch it upon only to the extent that it is
related to this aspect and do not discuss it on its own.)

It is a series of events and facts from Majorana's life that we now turn to.

\section{Travel from Palermo to Naples}

Here, the best known example is mysterious disappearance of Majorana.
Mainly, discussion of this point is based on comparison and estimate of
probabilities of different versions -- whether Majorana committed suicide,
retired to a cloister, left Italy for Argentina, etc. Meanwhile, even
cursory acquaintance with known circumstances forces us to conjecture that
the crucial point in the Majorana's plan was, in the first place, not this
or that concrete version but, rather, plurality of versions as such. (See
below in more detail.) After so long period we can hardly expect to find an
exact answer what precisely happened to Majorana -- in full accordance with
the Fermi's remark that "with his intelligence, once as he decided to
disappear or to make his body to disappear, Majorana would certainly have
succeeded" (\cite{amaldi}, p. 66). However, it seems we can understand
something in the mechanism itself chosen by Majorana for his purpose.

Let us remind basic points of this story according to the description done
in the Sciascia book \cite{it}. (We quote from the English translation of
this book which is published under the same cover with another Sciascia's
book \cite{eng}. For brevity, in what follows, we will simply give page's
numbers of \cite{eng} in parenthesis.) On 24 March of 1938 in evening
Majorana supposedly embarked at Palermo. The ship arrived at Naples on 25
March.

"The fact that Majorana had undertaken the return journey and landed in
Naples was confirmed by a return ticket which had been handed in and was
found at the head office of the "Tirrenia". The fact that a person who could
have been Ettore Majorana had traveled in the cabin assigned to him by that
ticket was confirmed by Professor Vittorio Strazzeri who had spent the night
in that cabin.

From the tickets handed in, it emerged that this cabin had been shared by
Charles Price, an Englishman, Vittorio Strazzeri and Ettore Majorana. It has
been impossible to trace Price. But there was no difficulty in finding
Professor Strazzeri, a lecturer at the University of Palermo.

To a letter from Ettore's brother (presumably accompanied by a photograph)
Professor Strazzeri replies that he doubts whether he did in fact travel
with Ettore Majorana and whether "the third man" was an Englishman. However
he's "absolutely convinced that "if the person who traveled with me was your
brother, he didn't commit suicide at least not before our arrival in Naples"
(163). As to the Englishman he questions the fact that he was called Price,
for he spoke Italian "like one of us southerners" and he had the somewhat
coarse manners of shopkeeper or even a more common person. This is indeed a
case of "the third man". But the problem is easily solved. Since Professor
Strazzeri exchanged a few words with the man he took for Charles Price and
none with the one he thought to be Ettore Majorana it's reasonable to
suppose that the man who didn't speak and whom Strazzeri later identified as
Ettore Majorana was in fact the Englishman, while the one who he was later
told was Price was a Sicilian, a Southerner, the shopkeeper he appeared to
be and who was travelling with Majorana's ticket. There would be nothing
surprising in that. Majorana could have gone to the Tirrenia' booking-office
at the right time and given his return ticket to a man who was about to buy
one and who may even have resembled him --- in age, stature, complexion
(it's not hard to find even among a handful of Sicilians one who is
typically Saracen)" (163).

The basic Sciascia's idea about substitutions looks reasonable but he did
not bring his thought to completion and did not draw conclusions which
inevitably arise from it. Apart from this, his version contains logical
inconsistencies or at least gaps. But, before proceeding further, we would
like to dwell upon the mistake made by the English translator in \cite{eng}.
The true meaning of the fragment "As to the Englishman he questions the fact
that he was called Price" from page 163 of \cite{eng} is opposite: Professor
Strazzeri does \textit{not }put in question that his room-mate's name was
Price but doubts that it was indeed an Englishman since that man spoke
Italian "like one of us southerners" and "had the somewhat coarse manners of
shopkeeper or even a more common person" (163). Italian original: "In quanto
all'inglese, non mette in dubbio che si chiamasse Price, ma parlava
Italiano"come noi, gente del sud" ed aveva modi piuttosto rozzi, da
negoziante o giu' di li' \cite{it}, page 60. (I thank Enzo Del Prete, Sergio
Caprara and Lanfranco Belloni for explanations concerning language problems.)

Let us now consider the situation with passengers in more detail. In the
room, beside Strazzeri, there have been two other passengers. For brevity,
let us denote them as A (a person with whom Strazzeri exchanged a few words
-- "Sicilian") and B (silent person). Let us suppose that, indeed (as
conjectured by Sciascia), A was fake Price whereas B was true Price. Hence,
it turns out that the passenger A introduced himself calling the name of the
passenger B but the passenger B did not protest. Then, it follows that
Majorana planned and realized an entire hoax due to agreement with both
passengers (not only with A, as was suggested by Sciascia). However, the
Sciascia's version is not the only possible one. Let us try to take into
account main variants, accumulating them in a table.

\begin{tabular}{|p{1.1cm}|p{6cm}|p{6cm}|}
\hline
& A & B \\ \hline
1 & Price & Passenger like Majorana under his name \\ \hline
2 & Real Price mimicking fake Price & Majorana \\ \hline
3 & Fake Price (not Majorana) & Real Price under the name of Majorana \\ 
\hline
4 & Fake Price making accent on his fictitious nature & Majorana as fake
Price of the "2nd order" \\ \hline
5 & Fake Price (not Majorana) & Fake Majorana (not Price) \\ \hline
6 & Majorana under the name of Price & Price under the name of Majorana \\ 
\hline
7 & Majorana under the name of Price & Fake Majorana (not Price) \\ \hline
8 & Real Price & Real Majorana \\ \hline
\end{tabular}

\bigskip

The straightforward variant of the hoax would have been restricted by
replacing Majorana by somebody under his name. But this is only particular
case 1 from a variety of versions. As a whole, the hoax under discussion
confirms without doubts the Fermi's remark cited above.

Now some explanations are in order. If manners and accent of the passenger A
were intentional (variants 2 and 4), it was made to aim investigators to a
wrong conclusion that Majorana was absent on the board, and the passenger B
was Price. It is just the version pushed forwarded by Sciascia who deemed
that "the problem is easily solved" (163). We did not include into the table
more artificial variants (for example, the variant in which A was real Price
and B was a fake Majorana). Variants 6 and 7 imply some minimum of player
capabilities in Majorana. We do not have such information but, nonetheless,
left these variants for completeness. However, in our view, what is the most
important, is not properties of concrete versions but, rather, the principal
aim at the plurality of variants. (For this reason we included variant 8:
although, as such, it does not contain direct substitutions, in a given
context it is not quite obvious and, therefore, looks as one of variants of
the hoax.) We deal with such a "performance" in which impossibility to find
an unambiguous solution is underlying meaning of the whole action rather
than a pragmatic task.

Sciascia pointed out that the passanger A had to bear resemblance with
Majorana (as in the course of investigation photos were to be showed only
after some period of time and the 3rd passenger did not try to remember the
appearance of room-mates intentionally, rather general resemblance was quite
sufficient for Majorana's purposes). Meanwhile, it follows from his
reasoning that it is the passenger B who had to bear resemblance with
Majorana to create false impression that Majorana was him. In our view, both
passengers -- A and B -- turn out to be potential substitutes of Majorana.
Correspondingly, they both (not only A, as Sciascia deemed) had to be like
Majorana. The passenger A probably combined apparent features of Majorana,
those of Sicilian (probably hired by Majorana) and of Englishman (name
"Price"). In a similar way, the situation was for B, but with the difference
that his reticence could be interpreted as an additional factor in favor of
identification with both Majorana and Englishman.

Thus, potential substitutions concerned all possible relationships within
the triad M -- A, M -- B, A -- B (here we denoted Majorana as M). In doing
so, the substitution was performed not as a replacement of one personage by
another with the entire set of his features but with entanglement of the
features themselves. As a result, instead of deterministic (though
incomplete, not known exactly) picture of the events, the essentially
probabilistic picture is obtained. It arises due to a kind of the exchange
effect and peculiar indistinguishability of constituents (in variant 6 of
the table such an effect occurs directly). Analogy with quantum mechanics is
obvious. Further, we will discuss it in more detail.

Now, we want to point out a rather amazing circumstance with translations of
the corresponding fragment that seems to be not incidental. In addition to
the mistake in the English version, there is also a misprint in the Russian
one. In the phrase "he spoke Italian "like one of us southerners" (163)
about the room-mate of Strazzeri, the Russian translator wrote "spoke
English" instead of "spoke Italian" \cite{rus}, page 280. (In Russian: "po
anglijski govoril "kak my juazhane"). Majorana has managed a kind of
joggling with names, languages, identities, so that "Englishman Price" spoke
Italian with Sicilian accent, etc. In such a situation it is not difficult
to go astray and mix two languages or forget, what was put in doubt - name
or nationality, etc. It can explain, why two qualified translators contrived
to make \ a mistake in the same short fragment. In our view, they can be
considered as victims of the Majorana's hoax. In this sense, the hoax has
been continuing and, so to say, in the double mistake of translators we can
distinct echo of Majorana's voice. Such relationships between text and life,
text and its author, text and audience forces us to remember writings of
Jorge Luis Borges.

Now we go on to another circumstances connected with Majorana's
disappearance.

\section{Letters}

At first, let us list a series of quotations from the Sciascia's book. "On
the evening of 25 March Ettore Majorana sailed on the 10:30 p.m.
Naples-Palermo mail-boat. He'd posted a letter to Carrelli, Head of the
Institute of Physics and had left one at his hotel addressed to his family.
His motives for not posting it are obvious: he'd reckoned how things should
and did turn out and he didn't want his family to get the news too brutally,
but by degrees. The letters have been read by many since Erasmo Recami, a
young physicist who is in charge of the Majorana documents at Domus
Galileiano, published them. But it seems worth reproducing them here. The
letter to Carrelli reads:

Dear Carrelli,

I've made a decision that was inevitable. There's not a single speck of
selfishness in it, but I do realize that my sudden disappearance will cause
some inconvenience to you and to my students. For this too I beg you to
forgive me, but more especially for having betrayed all the trust, true
friendship and sympathy you showed me during these months. Please remember
me to those I'd come to know and appreciate at the Institute, to Sciuti in
particular; of all these I shall preserve a fond memory at least until
eleven o'clock this evening, and perhaps beyond" (160).

The letter to family: "I have only one wish: do not wear black. If you must
conform to custom just wear some emblem of mourning, but not for more than
three days. After that remember me in your hearts, if you can, and forgive
me" (161).

"Carrelli hadn't yet received the letter addressed to him when he got an
urgent telegram from Majorana sent from Palermo, begging him to pay no
attention to it. (\ldots ) Later he received another letter from Ettore,
from Palermo, on paper bearing the heading "Grand Hotel Sole":

Dear Carrelli,

I hope you got my telegram and my letter at the same time. The sea rejected
me and I'll be back tomorrow at the Hotel Bologna travelling perhaps with
this letter. However I have the intention of giving up teaching. Don't think
I'm like an Ibsen heroine, because the case is different. I'm at your
disposal for further details" (161).

The actions of Majorana contain a number of contradictions. At first, he
sends a letter with a hint at suicide in preparation, later he abolishes it.
It would seem that a sender should hope the 2nd letter to come first to
cancel the 1st one with more than disturbed contents. However, instead of
it, he expresses hope that both letters arrived simultaneously, as if they
should demonstrate that mutually inconsistent version should be regarded on
equal footing. The letter to Carrelli, as Sciascia pointed out rightly,
contains an important ambiguity in the words "I shall preserve a fond memory
at least until eleven o'clock this evening, and perhaps beyond" (160). On
one hand, this can be understood as a possibility of denial of suicide, from
the other one -- as an uncertain possibility to retain memory already
"there".

"On 22 January he'd asked his mother to get his brother Luciano to withdraw
from the bank his own share of their joint account and to send it all to
him. And shortly before 25 March --- the day he left Palermo stating his
intention to commit suicide --- he withdrew his October to February salary
which, until then, he hadn't touched. He had no money-sense, as is obvious
from his neglecting for five months to cash his salary --- but that he
should cash it on the very eve of committing suicide is hardly credible. The
one simple explanation is that he needed it for what he was planning to do.

There is, of course, another less simple explanation: that the sheer
improbability of someone intending to commit suicide taking with him all the
money of which he disposed as well as his passport might consolidate his
mother's hopes that he hadn't killed himself and was still alive. But this
is invalidated by his request to the family not to wear mourning --- or if
they must, that it be some inconspicuous token, and only for three days (the
three days of the Sicilian 'strict mourning'). Clearly he wanted his death
to be taken for granted" (166).

In our view, from persistence with which Majorana creates all these
contradictions, a quite different conclusion follows. Majorana wanted to
achieve \textit{the existence on equal footing of different versions which
otherwise should be taken as mutually inconsistent -- whether he committed
suicide or survived}.

\section{Reference to Ibsen}

It is worth paying attention to one literature reference. In the letter to
Carrelli Majorana mentions Ibsen: "I am not young girl from one of Ibsen's
plays, you understand, the problem is much more great than that" (\cite%
{amaldi}, p.63) or "Don't think I'm like an Ibsen heroine, because the case
is different" (161). There are two Ibsens's plays in which a young girl or
woman commits suicide -- "The Wild Duck" and "Hedda Gabler" (161). Plots of
both are so far from the Majorana's situation that it may look strange why
Majorana mentions Ibsen at all. However, there is no any doubt that Majorana
thought over his own disappearance very carefully, so that his letters are
expected to contain no incidental or unnecessary details.

We can suggest the following explanation. In the given context what brings
the remark about Ibsen to the forefront is the fact that in Ibsen's works
suicides is encountered in more than one play. As a result, it becomes
impossible to \textit{identify the subject of suicide}. Thus, two different
motives -- suicide and ambiguity (deidentification) of personality overlap.

The last circumstance, as we will see, revealed itself also in other actions
of Majorana.

\section{Talk on the conference}

The following Majorana's escapade is known \cite{amaldi}, p. 32; \cite{pont}%
. When Fermi asked Majorana's permission to inform community about his
theory of nuclear forces at the Paris conference, Majorana agreed under
condition that his ideas must be ascribed to an old professor of electrical
engineering who had to attend the conference. Meanwhile, one can find some
inner meaning in this escapade if one takes into account that the subject of
discussion is the theory of exchange forces in which there exists some kind
of "exchange of essences" between particles composing a nucleus. The example
of such an "exchange effect" (although in one way only) had to happen in
case had the author of the Majorana's idea (in a sense, himself) would have
been replaced by someone else.

\section{How Majorana occupied position of professor}

"During four years -- from the summer of 1933 to that of 1937 -- he rarely
goes out and even more rarely turns up at the Institute of Physics. At a
given point he stops going there altogether" (154). But, suddenly, he sends
application to competition for the position of professor. Such
"socialization" looks rather unusual for Majorana, specifically against a
background of preceding period of isolated life. It is worth remembering
circumstances of the competition. "As usual the three winners had already
been tacitly selected before the competition took place: Gian-Carlo Wick,
first; Gulio Racah, second; Giovanni Gentille Junior, third" (157). The
result seemed to be determined in advance not only because of scientific
merits of Wick and Racah but also because farther of Giovanni Gentile had
some influence in ruling circles of the fascist regime. However, in case of
participation of Majorana in the competition there would be no doubt in his
win. Correspondingly, Gentile junior would not take a position at all. To
prevent such development of events, Majorana was appointed to the Chair of
Theoretical Physics of Naples "on the basis of his reputation and on the
strength of law instituted by the minister Casati and revived in 1935 by the
Fascists" (157). This gave possibility to complete the competition, as was
intended.

According to the Sciascia's opinion, Majorana "had only take part in the
competition as a bitter joke at the expense of his colleagues" (157). We
deem that the point is different. In case of success of his plan, Majorana 
\textit{would occupy other's} position in the fixed hierarchy. Apart from
this, inevitable further shift would occur in the arrangement of other
participants of the competition: the 1st would become the 2nd, the 2nd would
become the 3rd. Thus, some kind of the effect of deidentification would
occur, similar to that concerning the abortive talk about Majorana's theory
of nuclear forces (see above).

\section{Way of potential suicide}

Let us return to the problem of disappearance of Majorana. We tried to
substantiate that the hoax both as a whole and in details had to have
"probabilistic" nature and imitate probabilistic laws of quantum mechanics.
Then, it looks reasonable to extend this circumstance to the way of suicide
(real or fake) by itself -- death in sea \textit{waves} in such a way that
both alternative versions be possible, leaving uncertainty -- whether or not
suicide was committed. This gives an idea to link it to the principal role
of the\textit{\ wave} function in which the special probabilistic nature of
quantum mechanics (distinguishing it from the classical one) reveals itself.
Such an interpretation unities both motives -- probabilistic features and
wave character of the corresponding object. As a matter of fact, the
imitation of "waves of probability" in reality is obtained.

\section{Life, death and quantum mechanics: quantum version of Hamlet
question}

In the essay "Valore delle leggi statistiche nella fisica e nelle scienze
sociali" ("Role of statistical laws in physics and social sciences") written
during the period of isolated life in 1934 -- 1937, Majorana wrote: "A
radioactive atom's disintegration can force an automatic reactor to register
it with a mechanical effect made possible by adequate amplification. Thus
ordinary laboratory equipment is sufficient to prepare an extremely complex
and showy sequence of phenomena "set off" by the accidental disintegration
of a single radioactive atom. From a strictly scientific point of view there
is nothing to stop us from considering as plausible that an equally simple,
invisible and unpredictable vital phenomenon might be the cause of human
existence" (155).

In our view, this fragment should be juxtaposed with the problem, touched
upon by Schr\"{o}dinger in his seminal paper about paradox named later in
his honor as "paradox of the Schr\"{o}dinger cat". (This paper \cite{schr}
appeared in 1935 and we refrain from discussion whether or not Majorana read
it.) Schr\"{o}dinger wrote that in a closed space the event of decay of a
radioactive atom (which happens or does not happen with certain
probabilities according to the laws of quantum mechanics) can trigger a
chain of consequences resulting in death of a cat situated there. Thus, laws
of quantum mechanics are applied to macroscopic objects, life and death
forming a kind of superposition of different states. If our supposition is
correct, \textit{Majorana intended to model "superposition" of state of his
own life and death}. 

\section{From quantum mechanics to text of behavior}

Thus, the picture outlined above seems to allow composing more or less clear
fragment from what remained known about Majorana and circumstances of his
disappearance. According to explanations under discussion, Majorana
insistently imitated laws of quantum theory in surrounding world, his own
behavior and destiny -- the object of research merged with the subject.
Remembering that in quantum mechanics one cannot neglect the presence of
device with which a quantum object interacts in the course of measurement
process, one can note that in such merging the specific features of quantum
mechanics again manifested themselves. One can say that Majorana with the
ultimate honesty brought to the logical limit the relationship between
nature and an observer who is studying it, so important for quantum theory,
having included here his own life and death. We would like to stress once
again that concepts and phrases in everyday use of the kind "unity between
life and activity of a scientist" acquire in a given case quite non-trivial
meaning.

One reservation is in order. There is a crucial difference between the
imitation of quantum state described above and the true quantum state. The
object of an "experiment" (coinciding in this case with an
"experimentalist"), i.e. Majorana himself, was, of course, unambiguously
alive or dead in each moment of time -- quite another matter that his fate
after disappearance was (and remained) unknown. In this sense, the
opposition "life -- death" is of pure classical nature here. By contrast,
for a quantum object the probabilistic nature and co-existing of
alternatives (life and death in the case of the Schr\"{o}dinger cat) are
unavoidable in principle until the moment of measurement that selects one
among the set of alternatives. (For example, in the case of the Schr\"{o}%
dinger cat the counter reacts to the quantum decay of atom or does not react
and, correspondingly, the ampoule with poison is broken or not.) Thus,
possibilities of imitation of quantum properties (as well as, in essence, of
any imitation) were restricted but Majorana realized them in full measure.

Whatever unusual the features under discussion be, it is worthwhile to note
that parallels between human life and text were repeatedly noticed in
humanities (in quite different context, of course). Thus, Lotman wrote that
Pushkin intentionally permanently created his personality like an original
artistic work \cite{lot}. In the case of Majorana the feature in question
concerns destruction of life (at least for outer observers) rather than
construction. But, anyway, it was performed on so high level that this
forces us to recollect his scientific works that gives integrity to his
tragic fate.

\section{Majorana and Pirandello}

Composing features of human behavior into an united text can be connected,
in particular, with orientation to already existing texts. Among other
things, it concerns the situation of passing away. It is sufficient to
recall the role of the Goethe's novel "The Sorrows of Young Werther" that
provoked in Germany an epidemic of suicides. As far as Majorana is
concerned, we can point to a possible analogy between his disappearance and
the plot of L. Pirandello's novel "The Late Mattia Pascal". We recall that
the main character disappears from his world where he is believed to commit
suicide. Under a new name he starts a new life. However, some time later, he
imitates suicide under the second name and returns to his native environment.

This analogy is well known and became a common place in literature about
Majorana, it was even discussed in mass media (167). Nevertheless, in a
given context we would like to pay attention to some important nuances which
probably escaped from previous discussions. Direct motivation for this
hypothetic analogy consists in disappearance with imitation of suicide.
Meanwhile, it is essential that something more is contained here as well.
The transition between "this" and "that" worlds in the Pirandello novel
turns out to be two-way -- back and forth. Therefore, this could be
perceived by Majorana as equality of both states, literature version of the
superposition discussed above. Apart from this, such "quantum jumps" were
combined with the deidentification of personality since the character of
Pirandello changed his name twice (cf. what was said above about this
issue). In the given context it makes sense to recall also that one of
circumstances that promoted the choice of Mattia Pascal to leave his
previous life was his big loss in roulette, i.e. essentially probabilistic
factor.

It is also worth paying attention to the motif of doubling. In the obituary
devoted to Pascal his suicide was represented as repetition of the 1st
unsuccessful attempt from which he was allegedly saved by the guard. This
obituary notes that for the second time such a person (the name of the guard
is called) was missing. In other words, it discusses a "virtual story" with
a double, absence of which proved to be fatal for the character's fate.
Thus, not only a "real" character's fate but also its wrong version ascribed
to him, turns out to be connected with such a factor as plurality of
embodiments of the same personages (including the secondary one).
Respectively, one may speak about plurality of corresponding variants in the
individual history of a person.

Thus, a whole complex of motives actual for Majorana is revealed -- \textit{%
"transitions" between life and death, "transitions" between different
personalities, the role of probabilistic nature of the world in human's
fate, actuality of alternative variants}. For these reasons, we deem that
the given novel by Pirandello played even more important role in
disappearance of Majorana than one could expect. It concerns not only
borrowing from the plot but, rather, ideological and motif proximity --
Majorana found in the Pirandello works artistic interpretation of the
problems actual for him.

In turn, all this forces us to take seriously the potential role of one more
Pirandello's novel in the fate of Majorana. Discussing the analogy between
the behavior of Mattia Pascal and that of Majorana, Sciascia notes that "in
fact it conforms more to that of the hero of "Uno, nessuno e centomila"
(167) ("One, None, and a Hundred Thousand"). Unfortunately, Sciascia does
not explain his thought. Meanwhile, account for themes and motives discussed
above uncovers here striking resemblance between spiritual world of Majorana
and problems touched upon in the novel, connected with \textit{plurality of
personality}. For example, the author mentions a "hundred of thousands" of
Moscarda's (Vitangelo Moscarda is the hero of this novel), multitude of
heads which in fact constitute the same one, operates by multiple images of
the same reality, discusses the multitude of names for the same hero, etc.

In the novel, plurality of personality is connected with plurality of
viewpoints, i.e. a subjective property. As far as Majorana's world outlook,
there is a reason to expect that he proceeded much further, having tried to
turn for himself the property under discussion to the factor of objective
reality.

\section{On Majorana's motive forces}

Up to now, we mainly discussed formal features of the "text of behavior"
which was found in a number of Majorana's actions. It is quite natural to
ask how to give a meaningful interpretation to the structure found and, as
far as possible, to gain insight into the main Majorana's task connected
with this "text" and psychological motives of its "author". Here,
explanations will look more hypothetical than in the case of direct
systematization of separate observations. Nonetheless, we find such a way to
pose the problem not only quite rightful but also necessary -- as a matter
of fact, we should draw all possible conclusions from already uncovered
basic points.

As in the "text" under discussion fundamental issues of life and death were
touched upon in a quite unusual way, one can think that the corresponding
motive also had existential nature. Let us recall that Majorana "was a
pessimist by his nature and was permanently discontented by himself (and not
only by himself!") \cite{pont}, p. 27. With this circumstance taken into
account and generalizing our previous observations, it looks appropriate to
make the following conclusion. \textit{Majorana was not} \textit{content not
only with his own place and fate but with life and laws of existence as such}%
. In doing so, the main edge of dissatisfaction was pointed against
unambiguity and the absence of choice. On the individual level, it
manifested itself in the intention to ruin an unambiguous identity of life.
It is just the reason why he suggested ascribing his own ideas to somebody
else, tried to replace (for a future investigator of his disappearance)
himself by a whole set of persons, etc. On a more general level, this
revealed itself in the intention to cancel an unambiguous border between
life and death. It was not simply a question how to change psychological
attitude to such fundamental categories as life and death but, rather,
change their nature as such (at least, purely subjectively, as imitation).
In doing so, when Majorana planned and realized his so unusual
disappearance, he did not try to arrange the "performance" for spectators.
More exactly, it was him who was the only adequate spectator who understood
what and why he was doing -- all these substitutions on the ship seem to
have been acts of autocommunication.

If our psychological reconstruction is correct, it entails important
conclusions in what concerns the real fate of Majorana. There are three main
versions of what he did: 1) retired to a cloister, 2) committed suicide, 3)
hided himself in another country. In our view, now version 1) should be
certainly rejected. The concepts concerning existence and no-existence
described above were inconsistent with traditional Christian ones. And, to
the extent in which Majorana retains his religiosity, his concepts,
intentions and actions were certainly theomachistian, denying in the
existential rebellion basic laws established by the Creator. But, it seems
more probable that they simply had nothing to do with Christianity --
Majorana intended to raise for himself his own, individual world unlike
those known before.

Thus, only two versions remain. Plurality of personality, so important for
Majorana, gives reason for cautious optimism in the favor of version 3) --
actually, life in another country under an assumed name, would turn out to
be for him an analogue of another life. (See next Section for more detailed
discussion.) From another hand, Majorana could not miss the point that each
time he faced with only one reality but not two (or more) its alternative
versions simultaneously. With his permanent pessimism, such inevitable
disappointment could lead to a tragic result, so version 2) cannot be
excluded. Whatever real history of Majorana be after disappearance, in any
case we seem to be able to recover one important detail. In our view,
Majorana provided himself with an alternative identity card -- not only
because of necessity to solve a pragmatic task but also for the reasons
explained above. And, even if his life finished tragically, he had some time
to feel himself as someone else.

\section{Alternative worlds}

Generalizing previous observations, we must conclude that, for Majorana, the
most important property of the world was \textit{plurality of relaity}, the
existence of alternatives. As a result, in such an Universe possibilities
appear that would have been mutually inconsistent in a classical world.
Imaging other's reaction to his disappearance and the variants which they
had to take into account (thus, in a sense taking their viewpoint), Majorana
could himself turn into a conditional spectator, watching mentally different
version of his fate.

Then, such plurality means even something more than simply analogue or
imitation of laws of quantum mechanics. Even in quantum mechanics with its
unavoidable probabilistic nature in each experiment eventually only one
choice of alternatives occurs. Meanwhile, the preceding analysis forces us
to think that it was important for Majorana to embrace and feel different
alternatives just as \textit{real} events -- at least through perception of
other people. In other words, to live or feel different variants of his fate
including his own death. This circumstance strengthens arguments against
version 2) -- at least, Majorana could not commit suicide at once since he
needed to spend some time as an observer of different version of his own
fate. Thus, according to our approach, the key role in motive powers of
Majorana has been played by the idea about plurality of worlds, interpreted
not as a set of abstract possibilities (from which only one is realized),
but rather as \textit{real} variants.

It is striking that in recent years such an idea indeed appeared in science,
namely in quantum cosmology based on inflation theory and quantum theory.
According to ideas, pushed forward in \cite{vil1} - \cite{vil3}, there
exists an infinite number of universes but only a finite number of possible
histories. One of consequences consists in that if in a given region of
spacetime some history is realized, its other variants are inevitably
realized somewhere else. One may suppose that these ideas would turn out to
be congenial to Majorana. It is interesting that in recent years similar
ideas are becoming topical in art and literature. First of all, it concerns
the technique of "nonlinear narrative" due to which the same novel can have
different versions of the same events and different denouements. (One of the
brightest representatives of this direction is Milorad Pavich.)

In application to the Majorana case the paradigm under discussion means that
there exist worlds in which Majorana did commit suicide. However, there also
exist worlds in which Majorana has managed to overcome his pessimism and
survived. It remains to hope that the second variant is more frequent in
Universe.

\begin{acknowledgments}
The author is grateful to G. E. Gorelik for attentive reading the manuscript
and stimulating discussion.
\end{acknowledgments}

\end{document}